\documentclass[aip,apl,reprint]{revtex4-1} 

\usepackage{graphicx}
\usepackage{dcolumn}
\usepackage{bm}
\usepackage{amsmath, amssymb}
\usepackage{float}

\begin{document}

\title{Combined frequency and time domain measurements on injection-locked, constriction-based spin Hall nano-oscillators}

\author{T. Hache}
\affiliation{Helmholtz-Zentrum Dresden--Rossendorf, Institute of Ion Beam Physics and Materials Research, Bautzner Landstra\ss e 400, 01328 Dresden, Germany}
\affiliation{Institut f\"ur Physik, Technische Universit\"at Chemnitz, D-09107 Chemnitz}

\author{T. Weinhold}
\affiliation{Helmholtz-Zentrum Dresden--Rossendorf, Institute of Ion Beam Physics and Materials Research, Bautzner Landstra\ss e 400, 01328 Dresden, Germany}
\affiliation{Technische Universit\"at Dresden, 01062 Dresden, Germany}

\author{K. Schultheiss}
\affiliation{Helmholtz-Zentrum Dresden--Rossendorf, Institute of Ion Beam Physics and Materials Research, Bautzner Landstra\ss e 400, 01328 Dresden, Germany}

\author{J. Stigloher}
\affiliation{Experimentelle und Angewandte Physik, Universit\"at Regensburg, 93040 Regensburg, Germany}

\author{F. Vilsmeier}
\affiliation{Experimentelle und Angewandte Physik, Universit\"at Regensburg, 93040 Regensburg, Germany}

\author{C. Back}
\affiliation{Technische Universit\"at M\"unchen, 85748 Garching, Germany}

\author{S.S.P.K. Arekapudi}
\affiliation{Institut f\"ur Physik, Technische Universit\"at Chemnitz, D-09107 Chemnitz}

\author{O. Hellwig}
\affiliation{Helmholtz-Zentrum Dresden--Rossendorf, Institute of Ion Beam Physics and Materials Research, Bautzner Landstra\ss e 400, 01328 Dresden, Germany}
\affiliation{Institut f\"ur Physik, Technische Universit\"at Chemnitz, D-09107 Chemnitz}

\author{J. Fassbender}
\affiliation{Helmholtz-Zentrum Dresden--Rossendorf, Institute of Ion Beam Physics and Materials Research, Bautzner Landstra\ss e 400, 01328 Dresden, Germany}
\affiliation{Technische Universit\"at Dresden, 01062 Dresden, Germany}

\author{H. Schultheiss}
\affiliation{Helmholtz-Zentrum Dresden--Rossendorf, Institute of Ion Beam Physics and Materials Research, Bautzner Landstra\ss e 400, 01328 Dresden, Germany}
\affiliation{Technische Universit\"at Dresden, 01062 Dresden, Germany}

\date{\today}

\begin{abstract}


We demonstrate a combined frequency and time domain investigation of injection-locked, constriction-based spin Hall nano-oscillators by Brillouin light scattering (BLS) and time-resolved magneto-optical Kerr effect (TR-MOKE). This was achieved by applying an alternating current in the GHz regime in addition to the direct current which drives auto-oscillations in the constriction. In the frequency domain, we analyze the width of the locking range, the increase in intensity and reduction in linewidth as a function of the applied direct current. Then we show that the injection locking of the auto-oscillation allows for its investigation by TR-MOKE measurements, a stroboscopic technique that relies on a phase stable excitation, in this case given by the synchronisation to the microwave current. Field sweeps at different direct currents clearly demonstrate the impact of the spin current on the Kerr amplitude. Two-dimensional TR-MOKE and BLS maps show a strong localization of the auto-oscillation within the constriction, independent of the external locking.

\end{abstract}


\maketitle

Spin Hall nano-oscillators (SHNOs) are devices in which direct charge currents are converted into microwaves by driving large amplitude magnetic oscillations. Their main active component is a thin bilayer consisting of a heavy metal in direct contact with a ferromagnet. Due to the spin Hall effect (SHE)\cite{Hirsch1999,Ando2008, Hoffmann2013}, a charge current flowing in the heavy metal is converted into a transverse  spin current. This pure spin current can transfer angular momentum to the magnetic moments in the ferromagnetic layer via the spin transfer torque (STT) effect\cite{Slonczewski, Berger}. Depending on the direction of the charge current, this torque has a component either parallel or anti-parallel to the intrinsic Gilbert damping torque. Therefore, it can decrease or enhance the amplitude of magnetic oscillations, respectively. Above a certain threshold current the STT can fully compensate the intrinsic damping of the system and the magnetization starts to auto-oscillate in a free running state with frequencies in the GHz regime\cite{Slavin,Demidov2014,Demidov2014locking}. Compared to stacked spin torque oscillators\cite{Kiselev} which require charge currents flowing perpendicularly to the film plane, the simpler design of SHNOs makes them easier to fabricate, more robust, and allows the use of magnetic insulators which typically exhibit lower magnetic damping\cite{Collet2016}. Therefore, SHNOs are promising candidates as microwave or spin-wave sources for future communication technologies.

One of the issues of SHNOs is their limited coherence in the absence of external clocks for synchronisation. If the SHNO is in a free running state, the frequency of the oscillation is given by the geometry, material parameters, and external conditions, such as magnetic field, current, and temperature, causing fluctuations in phase and frequency \cite{Duan2014,Yang2015}. While measurements in the frequency domain have already shown that the coherence can be increased in arrays of SHNOs via mutual synchronisation \cite{Awad2016} or by locking to external microwave sources \cite{Demidov2014locking}, a direct comparison of spatially time- and frequency domain measurements is still missing. However, detecting the relative phase of coupled SHNOs in the time domain is a key requirement for optimizing the coherence and output power of larger arrays relevant for applications.
Here, we present a combined frequency and time domain study of free running and injection-locked SHNOs by Brillouin light scattering (BLS) microscopy and time-resolved magneto-optical Kerr effect (TR-MOKE) microscopy. This approach gives complementary information for the characterization of SHNOs which could not be obtained by one method alone. BLS measurements provide the full magnon spectrum independent from the actual locking state and can prove single-mode operation in the constriction while with TR-MOKE experiments strictly only detect the SHNO signal that is phase-correlated to the external stimulus giving further inside in the locking efficiency. 

\begin{figure}[b]
\begin{center}
\scalebox{1}{\includegraphics[width=8.5 cm, clip]{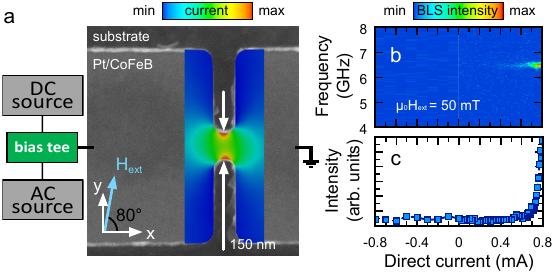}}
\caption{\label{fig1}(color online) (a) SEM image of the SHNO with a 150\,nm wide constriction. 
The overlay illustrates the calculated current density with blue (red)  representing lowest (highest) values, respectively. (b) BLS intensity detected between 4 and 8\,GHz as a function of the direct current ranging from $-0.8$ to $+0.8$\,mA. A magnetic field of $\mu_\mathrm{0}H_\mathrm{ext}=50\,\mathrm{mT}$ was applied. (c) Integrated BLS intensity as a function of the applied direct current. For currents above 0.7\,mA, a rapid increase in the intensity is observed, marking the threshold for the excitation of auto-oscillations. 
}
\end{center}
\end{figure}

Figure~\ref{fig1}a shows a scanning electron microscope (SEM) image of a  typical con\-stric\-tion-based SHNO. Using electron beam lithography and subsequent lift off, we patterned a Cr(1.5)/Pt(7)/Co$_{\mathrm {40}}$Fe$_{\mathrm {40}}$B$_{\mathrm {20}}$(5)/Cr(1.5) multilayer (all thicknesses in nanometers) into 1-$\mu$m-wide stripes that exhibit a sharp constriction with 150\,nm width and 110\,nm length in their center. For operating the SHNO, we apply direct and alternating electric currents using a bias-T. The resulting charge current flows along the $x$-direction as depicted in Fig.~\ref{fig1}a.

Due to the symmetry of the SHE, the STT induced damping-like torque can compensate the intrinsic damping most efficiently if the magnetization is aligned perpendicularly to the direction of the charge current. However, the injection locking results from a field-like torque caused by the dynamic Oersted field which is generated by the alternating current in the Pt layer. Since this field-like torque is perpendicular to the damping-like torque it can only affect the magnetization if it has a component parallel to the direction of the charge current. Therefore, we apply an external magnetic field $\textbf{H}_\mathrm{ext}$ at a slightly reduced angle of $\alpha=80^{\circ}$ relative to the direction of the {\it ac} and {\it dc} currents. This allows an effective compensation of the intrinsic damping via the STT and still enables the coupling to the dynamic magnetic field generated by the microwave current for synchronization.

The overlay in Fig.~\ref{fig1} presents the calculated direct current density\cite{calculation} which is highest inside the constriction due to the abrupt narrowing of the multilayer. Since the injected spin current is directly proportional to the current density in the Pt, the STT is maximal in the constriction which results in a well defined active region for auto-oscillations. 

The auto-oscillations were analyzed by BLS microscopy \cite{Sebastian2015}. This technique relies on the inelastic scattering of a monochromatic laser on the magnons and the subsequent frequency analysis of the scattered light using a high-resolution interferometer. The detected signal is directly proportional to the intensity of the magnetic oscillations. 

First, we characterize the free running SHNO, i.e., only driven by the applied direct currents. Figure~\ref{fig1}b shows the BLS frequency measured as a function of the applied {\it dc}. 
For the given direction of the magnetic field $\mu_\mathrm{0}H_\mathrm{ext}=50\,\mathrm{mT}$, we only observe magnetic oscillations for positive direct currents. At currents above 0.7\,mA, a clear signal emerges in the BLS measurements at frequencies around $f_\textrm{ao}=6.5$\,GHz. For the 150-nm wide constriction, only one auto-oscillatory mode  is effectively enhanced by the spin current. The onset of the auto-oscillation at 0.7\,mA can be even better seen in Fig.~\ref{fig1}c which plots the BLS intensity integrated over all frequencies. 

\begin{figure}[]
\begin{center}
\scalebox{1}{\includegraphics[width=7.8 cm, clip]{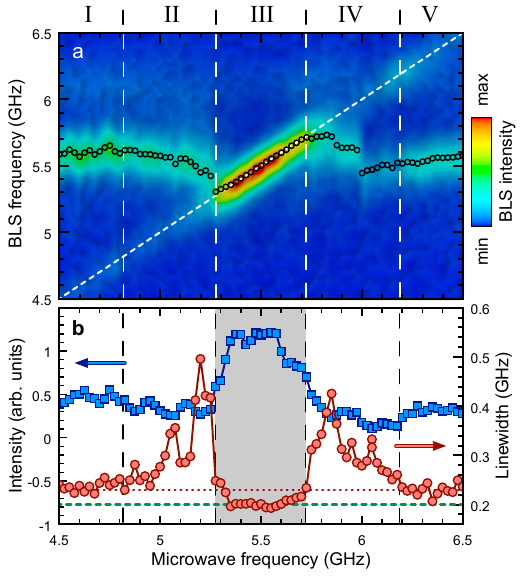}}
\caption{\label{fig2} (color online) (a) Locking characteristics of the SHNO measured as a function of the microwave frequency  for a fixed $I_\mathrm{dc}=1.5\,\mathrm{mA}$ and $\mu_\mathrm{0}H_\mathrm{ext}=50\,\mathrm{mT}$. The detected intensity is color coded with blue (red) representing minimum (maximum) values. Black dots mark the frequencies $f_\textrm{ao}$  of the excited auto-oscillations. Five different operating regimes can be identified: I,V free-running state of the auto-oscillation, II, IV frequency down- and up-pulling, respectively, and III the locked state. (b) Intensity (blue squares) and linewidth (red dots) of the auto-oscillation extracted from the measured BLS spectra.}

\end{center}
\end{figure}

We then demonstrate injection locking of this  {\it dc}-driven auto-oscillation to an additional  external microwave signal. Therefore, we kept the direct current fixed at  $I_\mathrm{dc}=1.5\,\mathrm{mA}$, well above the threshold for auto-oscillations. At $\mu_\mathrm{0}H_\mathrm{ext}=50\,\mathrm{mT}$, we recorded BLS spectra for different applied microwave frequencies $f_\textrm{ac}$ between 4.5 and 6.5\,GHz (Fig.~\ref{fig2}a). Please note that these measurements were done on another sample with the same nominal geometry. 

We identify several regions with distinct frequency evolutions which we label as regions I through V. In region I, up to $f_\textrm{ac}=4.9$\,GHz, the frequency of the auto-oscillation $f_\textrm{ao}$ stays constant around 5.6\,GHz, demonstrating that the auto-oscillation is only driven by the  {\it dc} current, i.e., in the free running state. At higher microwave frequencies $f_\textrm{ac}$ (region II, frequency down-pulling),  a redshift of $f_\textrm{ao}$ can be observed, indicating frequency pulling of the auto-oscillation. Starting at 5.3\,GHz (region III), the intensity of the auto-oscillation strongly increases and its frequency $f_\textrm{ao}$ follows the external trigger $f_\textrm{ac}$ (dashed, white line) within an effective locking range of approximately 440\,MHz. At even higher frequencies of the microwave signal, $f_\textrm{ao}$ slowly drops again accompanied by a decrease in intensity (region IV, frequency up-pulling) until the SHNO is again in the free running state (region V), solely driven by the direct current.

The increase of the auto-oscillation intensity within the locking range (region III) is quantified via fitting the BLS spectra with Lorentzian functions. The corresponding intensities and linewidths are plotted in Fig.~\ref{fig2}b. The increase in intensity of the locked auto-oscillation goes along with a decrease of its linewidth, indicating a higher degree of coherence. The dashed green line marks the average linewidth of 200\,MHz in the locking range whereas the dotted red line represents the average linewidth of 230\,MHz in the free running state. Note that the value given for the linewidth in the locked state already equals the spectral resolution of the BLS interferometer. Previous studies on similar geometries using microwave spectroscopy indicate that the actual linewidths of both the free running and the locked state are well below these values \cite{Demidov2014locking}. In regions II and IV, in which $f_\textrm{ao}$ is pulled down and up, respectively, the intensity of the oscillation is rather low and noisy which renders the fitting procedure less accurate and results in large variations of the extracted linewidths in regions II and IV.

\begin{figure}
\begin{center}
\scalebox{1}{\includegraphics[width=8.5 cm, clip]{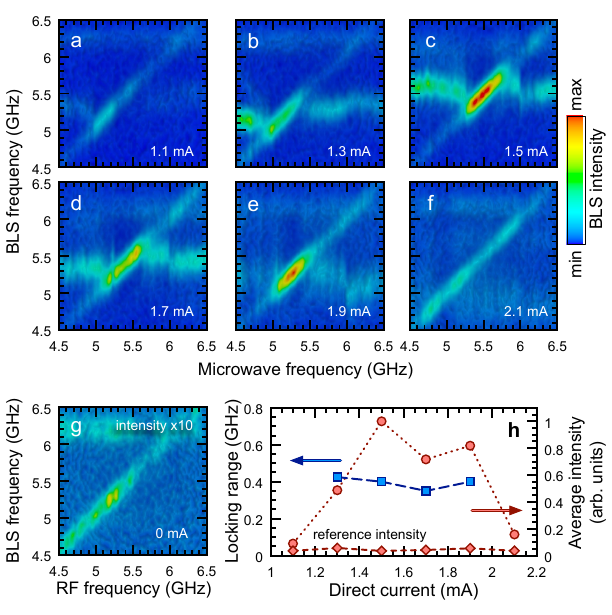}}
\caption{\label{fig3}(color online) (a)-(f) BLS spectra recorded as a function of the external microwave frequency for applied direct currents between 1.1 and 2.1\,mA. The detected intensity is color coded using the same scale for all measurements. (g) BLS measurement performed for the same range of microwave frequencies but without any  direct current applied. The overall intensity is multiplied by a factor of 10. (h) Locking range (blue squares) and average intensity (red dots) extracted from the data presented in a-f. Details on the reference intensity (red diamonds) are given in the text. All measurements were performed at $\mu_\mathrm{0}H_\mathrm{ext}=50\,\mathrm{mT}$ and an rf output of 12\,dBm.}
\end{center}
\end{figure}

As a next step, we investigated how the locking range and the average intensity of the measured signal depend on the applied  {\it dc}  and how it compares to  magnetic oscillations only driven by the microwave current. Figures~\ref{fig3}a-f show BLS measurements as a function of the external microwave frequency with direct currents ranging from 1.1 to 2.1\,mA. Note that Fig.~\ref{fig3}c repeats the data recorded at 1.5\,mA, which is  plotted in Fig.~\ref{fig2}a. All measurements were performed at the same microwave power of 12\,dBm and at a magnetic field of $\mu_\mathrm{0}H_\mathrm{ext}=50\,\mathrm{mT}$. The detected intensity is color coded using one identical scale for all graphs. Figure~\ref{fig3}g shows BLS spectra measured for the same kind of microwave sweep but without any applied  {\it dc} current. Even though the microwave power is identical to the other measurements, the response of the SHNO is rather weak. Therefore, we multiplied the intensities in Fig.~\ref{fig3}g by a factor of 10 in order to visualize the signal trace using the same intensity color scale. Note that the constant signal at a BLS frequency of approximately 6.2\,GHz is a laser mode. 

We draw several conclusions from this measurement series: $f_\textrm{ao}$ in the free running state peaks at 1.5\,mA and decreases both for lower and higher {\it dc} current.
We observe injection locking between direct currents of 1.3 and 1.9\,mA. Figure \ref{fig3}h shows the locking range (blue squares) and the average intensity within this interval (red dots) as a function of the direct current. While the intensity varies by a factor of two, the width of the locking range remains constant. The reference signal plotted as red diamonds in Fig.~\ref{fig3}h represents the intensity extracted from Fig.~\ref{fig3}g averaged within a microwave frequency interval that corresponds to the locking range for the respective direct current. Since neither the free running state nor the locked state were clearly observed for 1.1 or 2.1\,mA, we plotted the intensity averaged over the entire {\it ac} current frequency range. For those two current values there is no significant difference in the measured intensities no matter if there is a direct current applied during the {\it ac} current sweep or not. This is in strong contrast to the relative signal change that we observe in the measurements for which a locked state could be identified. From the comparison of the reference signal (diamonds) to the average intensity when both {\it dc} and {\it ac} curent are applied (dots), we conclude that a small signal due to the {\it ac} current can be effectively amplified by a factor of up to 20 with a spin current originating from the SHE.

\begin{figure}
\begin{center}
\scalebox{1}{\includegraphics[width=8.5 cm, clip]{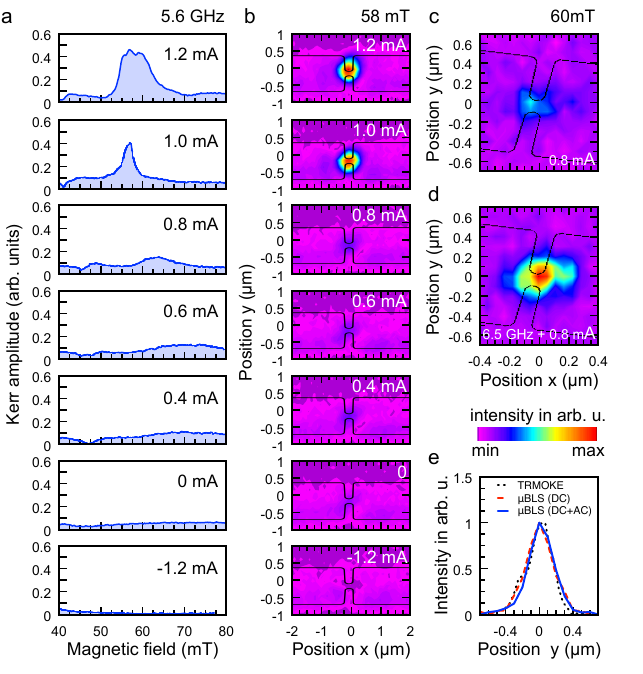}}
\caption{\label{fig4}(color online) (a) Kerr amplitude measured on the constriction  as a function of the magnetic field for a fixed applied microwave frequency of 5.6\,GHz and different direct currents between $-1.2$ and $+1.2$\,mA. The Kerr amplitude shows an emergent peak for positive currents starting at 0.8\,mA. 
(b) Two-dimensional TR-MOKE measurements of the SHNO recorded at $\mu_{0}\mathrm{H}_\mathrm{ext}=58\,$mT and ${f}_\mathrm{ac}=5.6\,$GHz. A clear localization of the auto-oscillation intensity within the constriction is observed. (c),(d) Two-dimensional BLS measurements of the auto-oscillation at 6.5\,GHz and $\mu_{0}\mathrm{H}_\mathrm{ext}=60\,$mT. In (c) only a direct current of 0.8\,mA was applied to drive the auto-oscillation, whereas (d) shows the intensity distribution of the SHNO locked at ${f}_\mathrm{ac}=6.5\,$GHz at a nominal power of 3 dBm. The locking leads to an overall increase of the intensity. (e) Intensity distributions extracted across the constriction from the TR-MOKE measurement at $\mathrm{I}_\mathrm{dc}=0.8\,$mA and both BLS measurements. }
\end{center}
\end{figure}

Based on this locking of the SHNO to the external microwave signal, it is also possible to detect auto-oscillations using stroboscopic techniques, such as TR-MOKE microscopy, which rely on the excitation of magnetization dynamics that are phase-stable  to an external trigger, i.e., the applied microwave current. To demonstrate this, we measured the Kerr amplitude at a fixed microwave frequency of 5.6\,GHz as a function of the external magnetic field  for different applied direct currents. As can be seen from the results in Fig.~\ref{fig4}a, no signals are detected for negative currents, in agreement with the BLS measurements. Instead, the Kerr signal is clearly suppressed below the level of zero current demonstrating the enhanced damping for negative currents. Only at a direct current of  0.8\,mA, a  peak emerges in the measured Kerr signal, indicating the onset of auto-oscillations, which rapidly grow in intensity with increasing current. 
Please note that the TR-MOKE measurements were performed on a different sample compared to the BLS measurements. Due to slight deviations in the lithographic fabrication and sputter process and due to the different amount of heat deposited by the BLS laser (532 nm) and the TR-MOKE laser (800 nm), small variations of the threshold currents and the locking characteristics are observed. 

To study the spatial distribution of the excited auto-oscillations, we performed two-dimensional TR-MOKE measurements for the same applied direct currents ranging from $-1.2$ to 1.2\,mA (Fig.~\ref{fig4}b). The magnetic field and the microwave frequency were fixed to 58\,mT and 5.6\,GHz, respectively.
For the two highest currents of 1.0 and 1.2\,mA, a clear localization of the locked auto-oscillation  within the constriction is observed. 

In order to analyze if the locking has any effect on this localization, we additionally mapped the two-dimensional intensity distributions of the free running and the locked SHNO using the BLS microscope. The external field and direct current were fixed to 60\,mT and 0.8\,mA, respectively. Figure~\ref{fig4}c shows the localization of the free running auto-oscillation at ${f}_\mathrm{ao}=6.5\,$GHz, whereas  Fig.~\ref{fig4}d plots the spatial profile of the auto-oscillation locked to an additional microwave current at ${f}_\mathrm{ac}=6.5\,$GHz. In the locked state,  an overall amplification of the auto-oscillation intensity by a factor of more than two is observed. Please note that the BLS measurements shown in Fig. 4c-e were done on the same sample as the measurements in Fig. 1b,c.

However, the spatial confinement within the constriction is not changed when the auto-oscillation is locked. To better visualize this, we compare the normalized intensity detected across the constriction in the TR-MOKE measurement at 0.8\,mA and both BLS measurements in Fig.~\ref{fig4}e.  There is no significant difference in the widths of the three  localization profiles. Gaussian fits of the data result in full width half maxima  of about 350\,nm for the free running and locked auto-oscillation measured by BLS as well as the locked auto-oscillation measured by TR-MOKE. Note, that the detected localization profiles are a convolution of the laser spot with the actual width of the auto-oscillation, which results in profiles much broader than the 150\,nm width of the constriction.

In conclusion, we experimentally demonstrate that injection locking of constriction-based SHNOs enables time-domain investigations using TR-MOKE. It is shown via BLS measurements that within a certain locking range, the frequency of the auto-oscillation can be forced to the value of an external stimulus which goes along with an increase in intensity and a reduction in linewidth. Even more importantly, the injected spin current yields an amplification of the {\it ac} signal by a factor up to 20. Ultimately, the coherent locking of the auto-oscillation allowes us to employ TR-MOKE to study the SHNOs. Spatially resolved BLS and TR-MOKE measurements prove that the excited auto-oscillation is strongly localized inside the constriction, independent if it is locked or in the free running state. We believe that our results show the feasibility of time-domain measurements on injection-locked auto-oscillations which  in future will allow to analyze and, therefore, optimize the phase relations inside arrays of SHNOs for further increasing their output power.

\begin{acknowledgments}
Financial support by the Deutsche Forschungsgemeinschaft is gratefully acknowledged within program SCHU2922/1-1.  K.S. acknowledges funding within the Helmholtz Postdoc Programme. Samples were fabricated at the Nanofabrication Facilities (NanoFaRo) at the Institute of Ion Beam Physics and Materials Research at HZDR. 
\end{acknowledgments}


\end{document}